\documentclass[aps,pra,superscriptaddress,address,amsmath,amssymb,showpacs]{revtex4}

\usepackage{graphicx}
\usepackage{graphicx} 
\usepackage{amsmath} 
\usepackage{amssymb}
\usepackage{hyperref}
\usepackage[usenames]{color} 
\newcommand{\beq}{\begin{equation}}
\newcommand{\enq}{\end{equation}}
\newcommand{\bes}{\begin{split}}
\newcommand{\ens}{\end{split}}

\begin{document}


\title{Topological surface flat bands in optical checkerboard-like lattices}

\author{Tomi Paananen and Thomas Dahm}

\affiliation{Universit\"at Bielefeld, Fakult\"at f\"ur Physik, 
Postfach 100131, D-33501 Bielefeld, Germany}

\date{\today}

\begin{abstract}
We present comparatively simple two-dimensional and 
three-dimensional checkerboard-like optical lattices possessing nontrivial
topological properties 
accompanied by topological surface states. 
By simple tuning of the parameters
these lattices can have a topological insulating phase,
a topological semi-metallic phase, or a trivial insulating phase.
This allows study of different topological phase transitions
within a single cold atom system. In the topologically nontrivial
phases flat bands appear at the surfaces of the system.
These surface states possess short localization lengths such that
they are observable even in systems with small lattice dimensions.

\end{abstract}



\maketitle

\section{Introduction}

The theoretical prediction \cite{Bernevig,Fu} and experimental discovery of topological
insulators \cite{Koenig,Hsieh1} has spurred the interest in nontrivial topological phases.
The common property of these systems is the fact that surface states
are protected by topological quantum numbers, making them particularly
stable against different kinds of perturbations \cite{Hasan}.
In particular, topological semi-metals like Dirac semi-metals and Weyl
semi-metals and their unusual surface states have been studied recently
in condensed-matter systems \cite{Wan,Burkov,ZKLiu,Borisenko,Jeon,Imada}. 

Optical lattices with cold atoms are perfect tools to simulate condensed
matter problems. Using optical lattices one can modify lattice depths and
lattice structures \cite{Petsas1994a,Orzel2001a,Chin2006a,Windpassinger}. 
That is why almost any condensed matter system can be simulated using optical lattices.
Interactions between cold atoms can also be tuned via the 
Feshbach resonance~\cite{Chinreview,Bartenstein2005a}. 
Several proposals have been made to realize topologically nontrivial
states in cold atom systems \cite{Stanescu,Sun,Lan,Buchhold,Mei,Klinovaja,Kennedy,Grusdt}
and recently first experimental realizations in 
optical lattices have been demonstrated \cite{Atala,Jotzu,Aidelsburger}.

In the present work we make a specific proposal, how topological surface flat
bands can be realized using comparatively simple optical lattices.
Flat band states are particularly interesting, because the group
velocity vanishes, and highly localized states can be formed.
Also, the effect of interactions becomes particularly important in flat bands
allowing new states of quantum matter \cite{Wu,Weeks,Sun2} 
or interaction
driven phase transitions, like for example a ferromagnetic state in
graphene nanoribbons \cite{Yazyev} or surface superconductivity
with high critical temperature~\cite{Kopnin2011,Kopnin2012}.
A distinction has to be made between bulk flat bands that appear
through the whole system in certain types of lattices
\cite{Wu,Weeks,Sun2,Bercioux,Apaja,Guzman,Chern}
and surface flat bands that are guaranteed to exist at the surface of a
topologically nontrivial system as a consequence of bulk-boundary
correspondence \cite{RyuHatsugai,Matsuura,Paananen_1,Paananen_2}. 
The present work is concerned with the latter case. Such kind of
topological surface flat bands have been found 
previously in other condensed matter systems like graphene,
superfluid $^3$He, or unconventional superconductors 
\cite{Nakada,Machida,Silaev,SchnyderTimmPRL,BrydonNJP,PALee,Tewari,Lau,Hu,Tanaka,Kashiwaya,Golovik,SatoTanaka1,SatoTanaka2,Volovik,Assaad,Roy}
and may also appear in topological insulators with a time-reversal breaking ferromagnetic exchange
field~\cite{Paananen_1,Paananen_2,Goette}.
The appearance of flat bands in $d$-wave superconductors
as surface Andreev bound states has been studied intensively in the past both
theoretically and experimentally~\cite{Hu,Tanaka,Kashiwaya,RyuHatsugai,Sato,Fogelstroem,Walter,Aprili,Krupke,Iniotakisprb,Chesca1,Chesca2,Iniotakis,Graser,Zare,Zhuvarel,Dahm}.
Using optical lattices
with ultra-cold atoms such surface flat bands and in particular
the influence of interactions on them could be studied in a very controlled way.
Experimentally, surface states in cold atom systems can be detected
by Bragg spectroscopy \cite{Sun,Buchhold} or a combination of Ramsey
interference and Bloch oscillations \cite{Abanin}.

In this work, we will present a two dimensional (2D) and a three dimensional (3D) optical lattice model,
possessing one dimensional and two dimensional surface flat bands, respectively.
We will give simple analytical criteria for existence and location of these flat bands.
Using two independent means - exact numerical diagonalization and an
analytical method - we demonstrate that the system can be tuned from a topological insulating phase via
a topological semi-metallic phase to a trivial insulating phase by tuning the intensity of
the lasers creating the lattice. This allows study of various interesting
topological phase transitions within a single model.
We also show that in the 3D case the flat bands are always two dimensional, and the flat bands can be
doubly degenerate under some circumstances. Flat bands
can appear both for an insulating as well as a semi-metallic bulk phase.
The appearance of the flat bands can be understood in terms of a classification recently
proposed by Matsuura et al~\cite{Matsuura} using a topological
invariant in the presence of a chiral symmetry.

\section{Models}

\subsection{Two-dimensional model}
\begin{figure}[t]
 \includegraphics{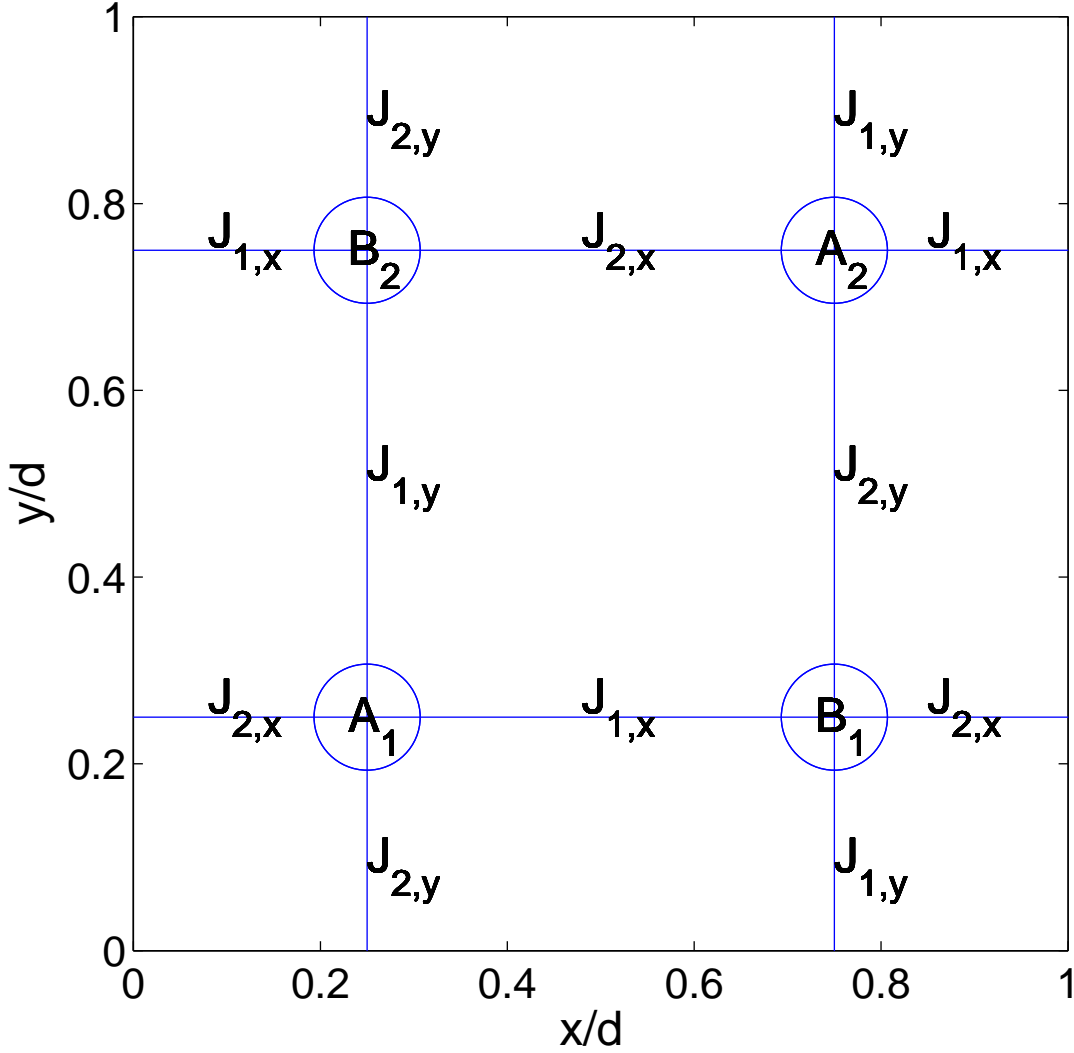} 
\caption{\label{Fig1}
(Color online) Schematic figure of the sublattice structure and the corresponding hopping strengths.
}
\end{figure}

Our model is a tight binding checkerboard model with different forward and backward hoppings.
The Hamiltonian can be written as
\beq
\label{eq:Ham_2D_1}
\begin{split}
\hat H_{2D}&=\sum_{m,n}\bigg[-J_{1,x}\hat a_{1,m,n}^{\dagger}\hat b_{1,m,n}-J_{2,x}\hat a_{1,m,n}^{\dagger}\hat b_{1,m-1,n}
-J_{1,x}\hat a_{2,m,n}^{\dagger}\hat b_{2,m+1,n}-J_{2,x}\hat a_{2,m,n}^{\dagger}\hat b_{2,m,n}\\
&-J_{1,y}\hat a_{1,m,n}^{\dagger}\hat b_{2,m,n}-J_{2,y}\hat a_{1,m,n}^{\dagger}\hat b_{2,m,n-1}
-J_{1,y}\hat a_{2,m,n}^{\dagger}\hat b_{1,m,n+1}-J_{2,y}\hat a_{2,m,n}^{\dagger}\hat b_{1,m,n}+h.c.\bigg],
\end{split}
\enq
where $m$ and $n$ are the unit cell indices, and $\hat a$ and $\hat b$ are
annihilation operators
for different sublattice sites. This Hamiltonian
is essentially a two-dimensional generalization of the dimerized optical lattice that has
been studied in Ref.~\onlinecite{Atala}. It consists of four sublattice sites
per unit cell, as shown in Fig.~\ref{Fig1}.
If the periodicity of the lattice is $d$, the lattice spacing of the sublattice is $d/2$. 
We can assume without loss of generality that $J_{1,\alpha}\geq J_{2,\alpha}$
(if this is not the case we can always 
relabel the hopping strengths and the sublattice sites). 
In Appendix A we discuss how such an optical lattice can be created by a
certain laser arrangement. Recently there has been intensive experimental
effort to create either spin-orbit coupling \cite{Kennedy} or non-Abelian
gauge potentials \cite{Osterloh,Ruseckas}
in cold atom systems in a desire to simulate topological insulators.
We note, that neither spin-orbit coupling nor
a non-Abelian gauge potential is needed to create the present lattice.
Nevertheless, topological surface states appear for certain parameter ranges,
as shown below.

If one takes the Fourier transform of the Hamiltonian, it can be written as
\beq
\label{eq:Ham_2D_2}
H_{2D}=\sum_{\bf k}\hat \Psi^{\dagger}_{\bf k}
\begin{pmatrix}
0 & 0 & f^*_x(k_x) & f^*_y(k_y)\\
0 & 0 & f^*_y(k_y) & f^*_x(k_x)\\
f_x(k_x) & f_y(k_y) & 0 & 0\\
f_y(k_y) & f_x(k_x) & 0 & 0
\end{pmatrix}\hat \Psi_{\bf k},
\enq
where ${\bf k}=(k_x,k_y)$,
\[
\hat \Psi_{\bf k}=(\hat a_{1,\bf k},\hat a_{2,\bf k},\hat b_{1,\bf k},\hat b_{2,\bf k})^T,
\]
and
\beq
\label{eq:k_functions}
f_{\alpha}(k_{\alpha})=-J_{1,\alpha}e^{ik_{\alpha}d/2}-J_{2,\alpha}e^{-ik_{\alpha}d/2}.
\enq
The Hamiltonian possesses particle-hole symmetry, if the system is half-filled
with two fermions per unit cell. Also, time-reversal symmetry, parity
symmetry, and most importantly a chiral symmetry 
\beq
\label{eq:S}
S= \begin{pmatrix}
1 & 0 & 0 & 0 \\
0 & 1 & 0 & 0 \\
0 & 0 & -1 & 0\\
0 & 0 & 0 & -1
\end{pmatrix}
\enq
is respected, i.e. $H_{2D}$ and $S$ anticommute. As has been discussed in Ref.~\onlinecite{Schnyderpt,Ryu},
the chiral symmetry $S$ is essential for the existence of edge flat bands.
Note that the phases of $f_x$ and $f_y$ cannot be transformed away
by a gauge transformation as they correspond to a Berry phase \cite{Zak}.

The single particle Hamiltonian has an off-diagonal block form, thus the
methods from Ref.~\onlinecite{Schnyder,Matsuura} can
be used to treat the Hamiltonian. In our case the block is given by
\beq
\label{eq:Block_2d}
D_{2D}({\bf k})=
\begin{pmatrix}
f_x(k_x) & f_y(k_y)\\
f_y(k_y) & f_x(k_x)
\end{pmatrix}
\enq
For a boundary perpendicular to the $x$-direction, the existence of edge flat
bands is connected to the value of the following winding number~\cite{Schnyder,Matsuura}
\beq
\label{eq:sch_integral_2D_1}
w(k_y)=\frac{1}{2\pi}\text{Im}\left(\int_{-\pi}^{\pi}\,dk_x\, \frac{\partial_{k_x}\text{Det}(D_{2D}({\bf k}))}
{\text{Det}(D_{2D}({\bf k}))}\right).
\enq
Here we have set $d=1$. We can define the path $\gamma(k_x)=\text{Det}(D_{2D}({\bf
  k}))=f_x(k_x)^2-f_y(k_y)^2$ in the complex plane. Then, 
this integral can be written as a path integral as follows
\beq
\label{eq:sch_integral_2D_2}
w(k_y)=\frac{1}{2\pi i}\oint_{\gamma} \frac{dz}
{z}.
\enq
This shows that the winding number $w(k_y)$ is always an integer. If it equals
to zero we do not have a flat band at the surface, otherwise we have.
However, this formula is not very useful to determine the values of $k_y$ for
which a flat band exists.
Instead, setting $z=e^{ik_{x}d}$ integral~\eqref{eq:sch_integral_2D_1} can be
mapped to a path integral over the unit circle, and the integrand is then given by
\beq
\label{eq:sch_integrand_2D_1}
f_{2D}(z)=\frac{J^2_{1,x}z^2-J^2_{2,x}}{z[(J_{1,x}z+J_{2,x})^2-zf_y(k_y)^2]}=\frac{-1}{z}+\frac{1}{z-w_{+}}+\frac{1}{z-w_{-}},
\enq
where
\beq
\label{eq:Roots_2D}
w_{\pm}=\left(\frac{f_y(k_y)\pm\sqrt{f_y(k_y)^2-4J_{1,x}J_{2,x}}}{2J_{1,x}}\right)^2.
\enq
Let us assume first that $J_{1,x}>J_{2,x}$.
The absolute values of $w_{\pm}$ determine whether we have a flat band or not.
It is easy to prove that one of the absolute values is always smaller 
than $J_{2,x}/J_{1,x}<1$. Thus, if both are smaller than 1 we have a flat
band, if not we do not have a flat
band. It turns out that
\beq
\label{eq:max_a_value_2D}
\begin{split}
&\max(|w_{+}|,|w_{-}|)\leq \left(\frac{(J_{1,x}-J_{2,x})|V_{k_y}|+\sqrt{[(J_{1,x}-J_{2,x})|V_{k_y}|]^2+4J_{1,x}J_{2,x}}}{2J_{1,x}}\right)^2
,\text{if}\,\, |V_{k_y}|<1\\
&\max(|w_{+}|,|w_{-}|)=1,\hspace{252pt} \text{if}\,\, |V_{k_y}|=1\\
&\max(|w_{+}|,|w_{-}|)\geq \left(\frac{(J_{1,x}-J_{2,x})|V_{k_y}|+\sqrt{[(J_{1,x}-J_{2,x})|V_{k_y}|]^2+4J_{1,x}J_{2,x}}}{2J_{1,x}}\right)^2
,\text{if}\,\, |V_{k_y}|> 1,
\end{split}
\enq
where
\[
|V_{k_y}|^2=\left(\frac{J_{1,y}+J_{2,y}}{J_{1,x}+J_{2,x}}\right)^2\cos^2(k_y/2)
+\left(\frac{J_{1,y}-J_{2,y}}{J_{1,x}-J_{2,x}}\right)^2\sin^2(k_y/2)=
a_y^2\cos^2(k_y/2)+b_y^2\sin^2(k_y/2)
\]
with $a_y=\left| \frac{J_{1,y}+J_{2,y}}{J_{1,x}+J_{2,x}} \right|$ and
$b_y=\left| \frac{J_{1,y}-J_{2,y}}{J_{1,x}-J_{2,x}} \right|$.
Equation~\eqref{eq:max_a_value_2D} shows that $\max(|w_{+}|,|w_{-}|)<1$, if $|V_{k_y}|<1$, and $\max(|w_{+}|,|w_{-}|)>1$, if $|V_{k_y}|>1$.
From this we can deduce that we have a flat band only if $|V_{k_y}|<1$.
From equation~\eqref{eq:max_a_value_2D} we can deduce the following:
If both $a_y<1$ and $b_y<1$, we have an edge flat band for all $k_y$, and the
bulk is an insulator. 
Thus in this case the flat bands are isolated.
If $a_y\geq 1$ and $b_y<1$, we have a flat band for 
\[
|k_y|>2\arccos\left(\sqrt{\frac{1-b_y^2}{a_y^2-b_y^2}}\right),
\]
and the bulk is a topological semi-metal.
If $a_y<1$ and $b_y\geq 1$, we have a flat band for 
\[
|k_y|<2\arccos\left(\sqrt{\frac{b_y^2-1}{b_y^2-a_y^2}}\right),
\]
with the bulk being a semi-metal, too.
If both $a_y\geq 1$ and $b_y\geq 1$, we have no flat bands. If $a_y>1$ and
$b_y>1$, we have an insulating bulk without edge states, i.e. a topologically
trivial insulator. Thus, by tuning the optical lattice potential via the
parameters $a_y$ and $b_y$ it becomes
possible to drive the system into different topological phases. 

\begin{figure}[t]
\begin{tabular}{ll}
 \includegraphics[width=0.49 \columnwidth]{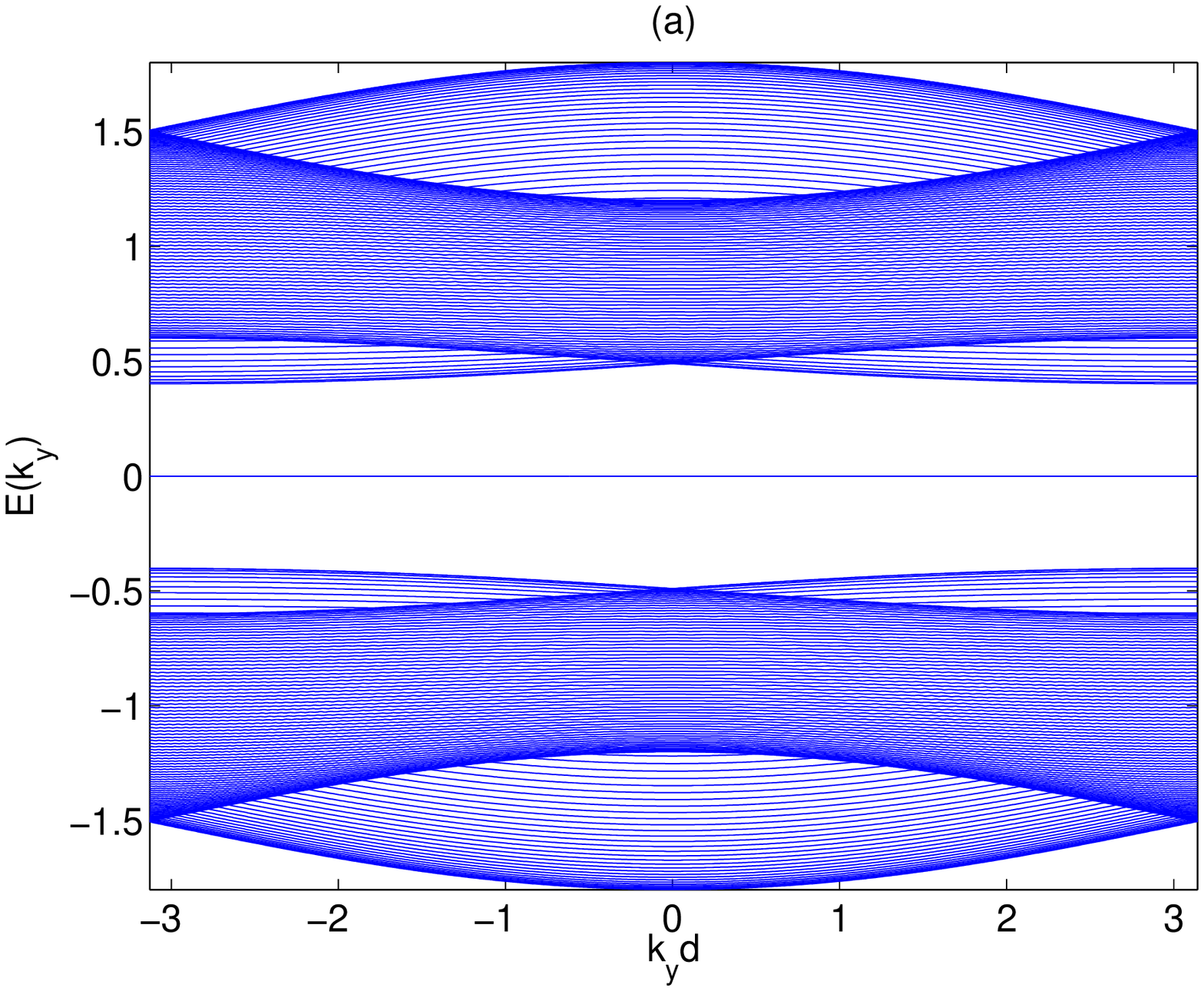} & \includegraphics[width=0.49 \columnwidth]{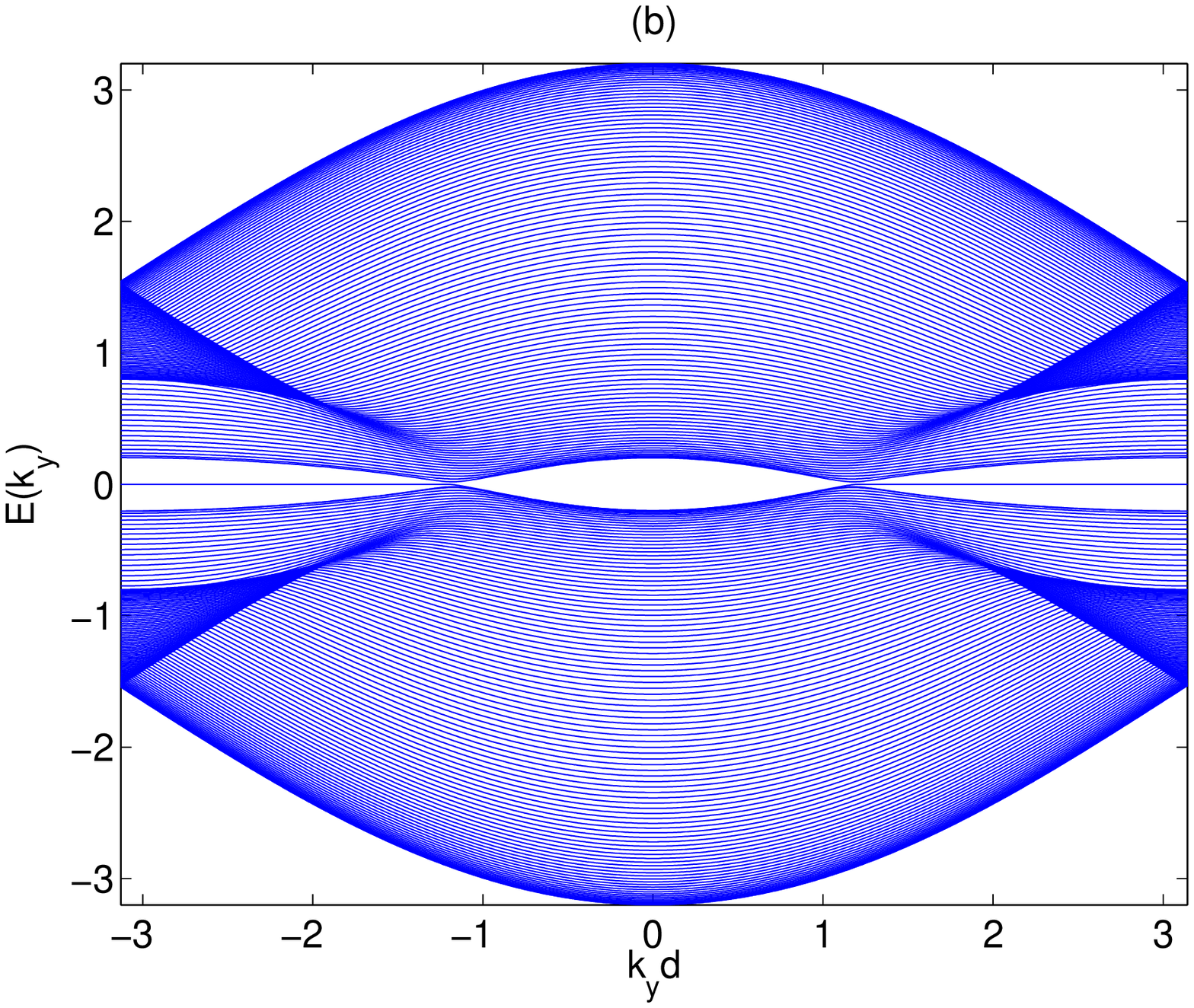}\\
 \includegraphics[width=0.49 \columnwidth]{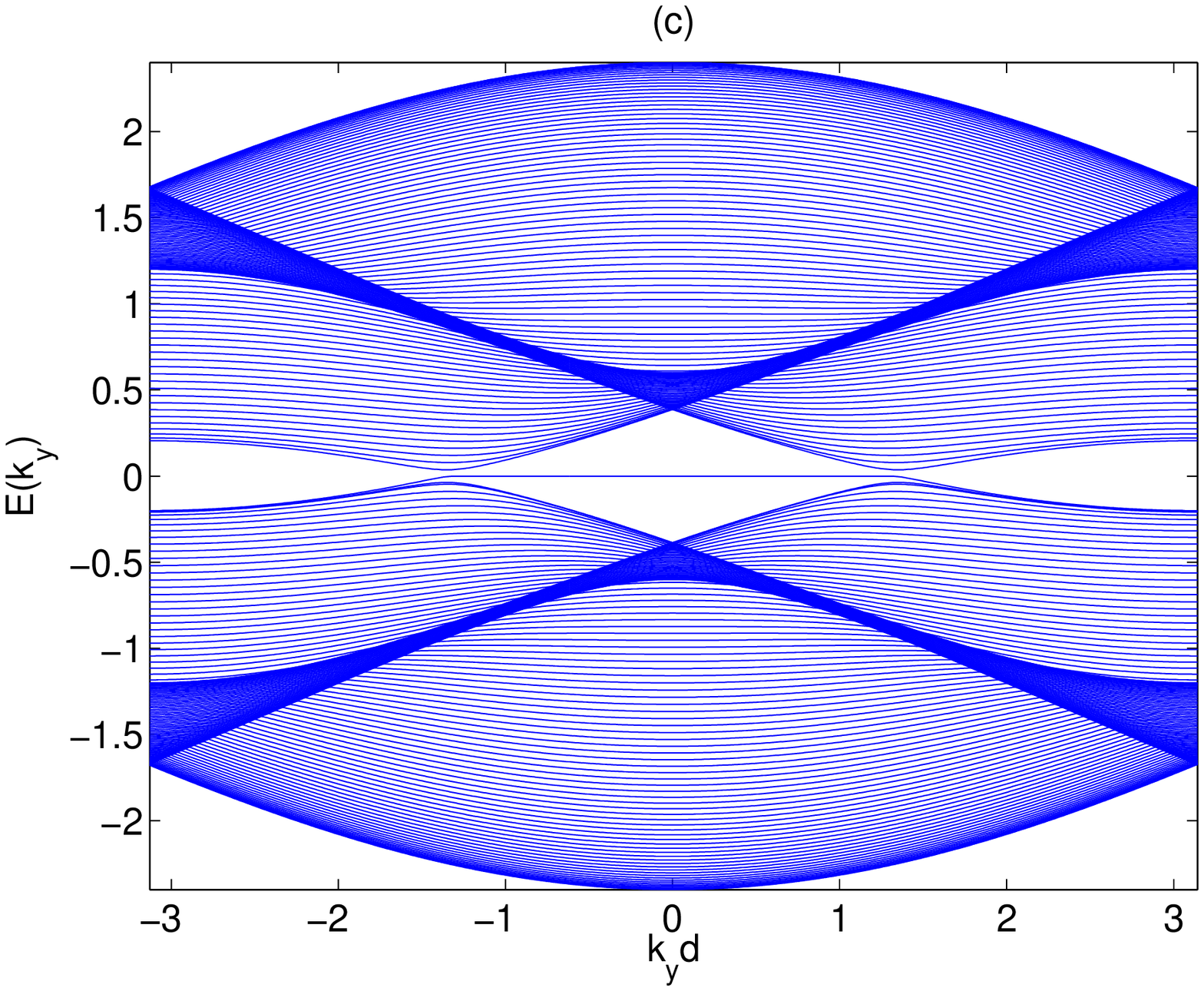} & \includegraphics[width=0.49 \columnwidth]{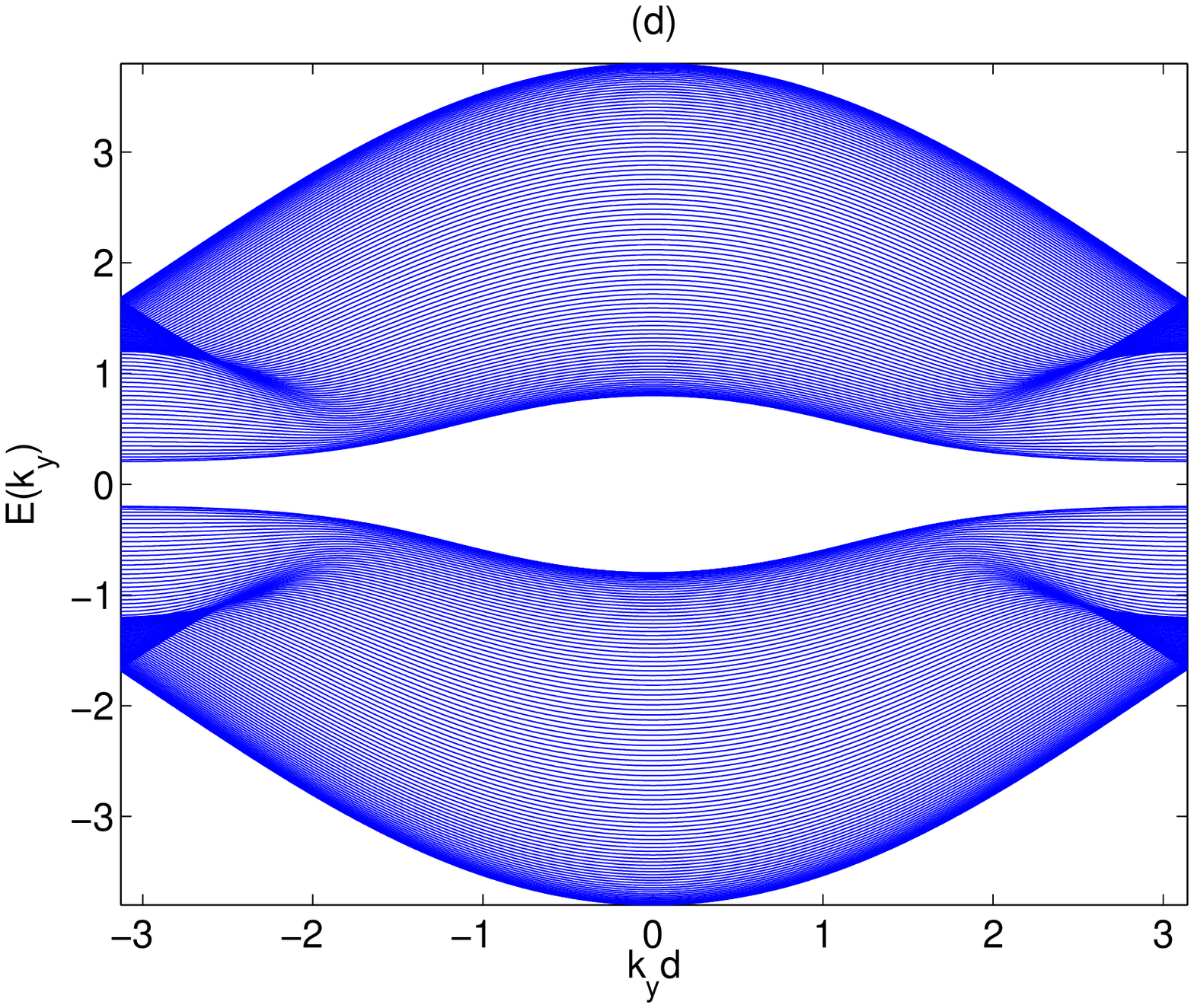}
\end{tabular}
\caption{\label{Fig2}
(Color online) Numerical 2D dispersions of bulk and edge states for $J_{1,x}=1.0$ and $J_{2,x}=0.5$ with different values of
$J_{1,y}$ and $J_{2,y}$. In figure (a) $J_{1,y}=0.2$ and $J_{2,y}=0.1$, in figure (b) $J_{1,y}=1.0$ and $J_{2,y}=0.7$,
in figure (c) $J_{1,y}=0.8$ and $J_{2,y}=0.1$, and in figure (c) $J_{1,y}=1.5$ and $J_{2,y}=0.8$.}
\end{figure}

\begin{figure}[t]
\begin{tabular}{ll}
 \includegraphics[width=0.49 \columnwidth]{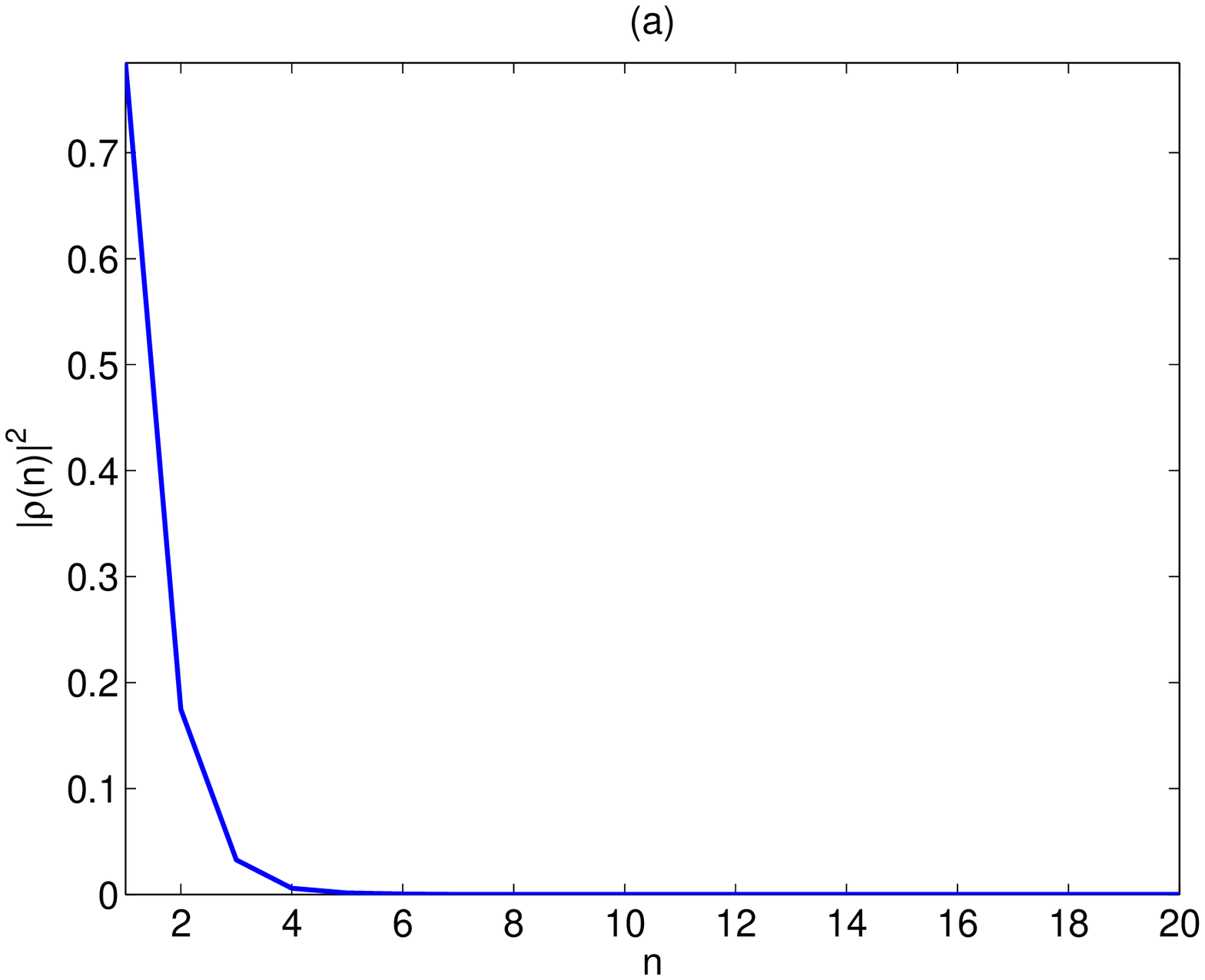} & \includegraphics[width=0.49 \columnwidth]{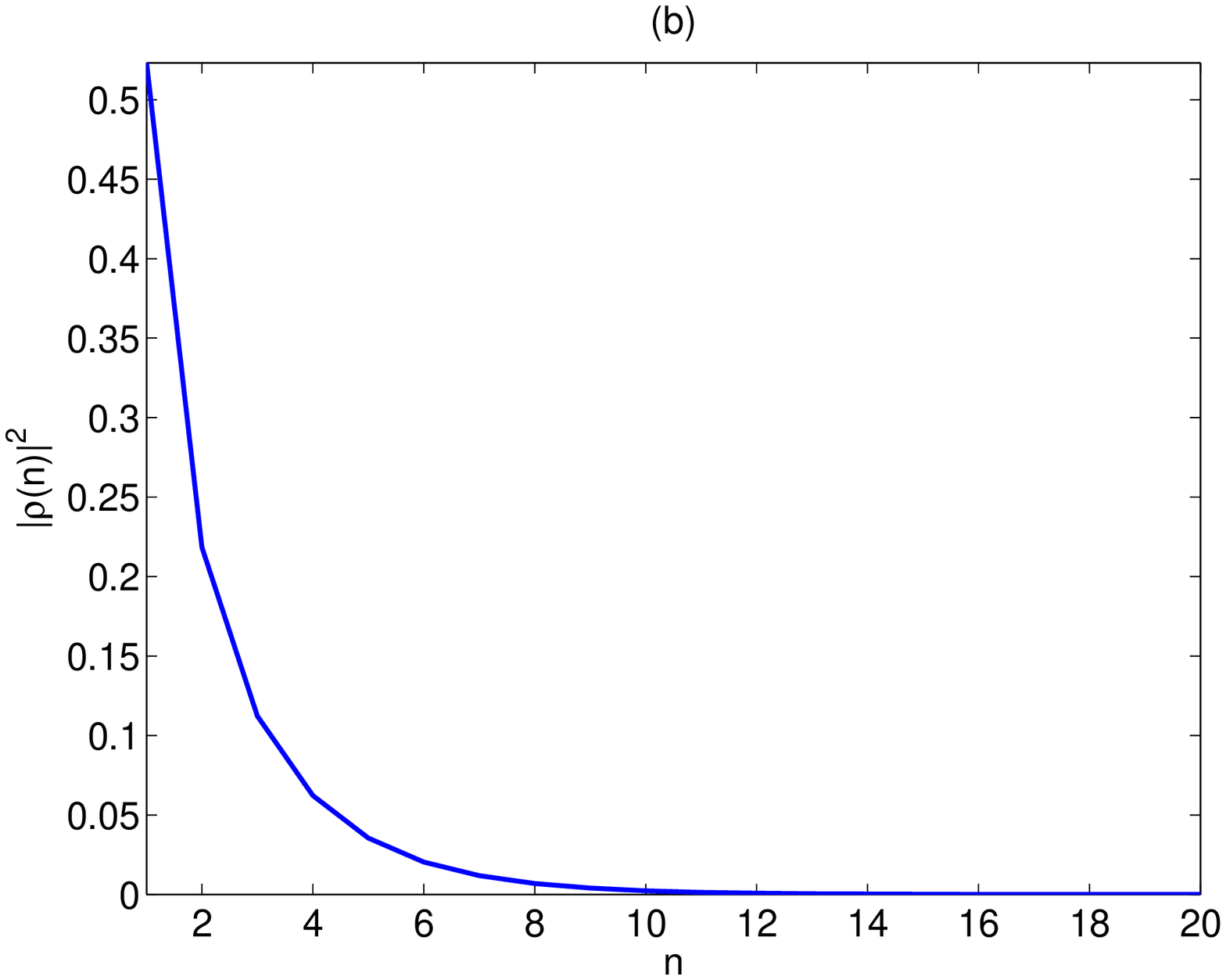}\\
 \includegraphics[width=0.49 \columnwidth]{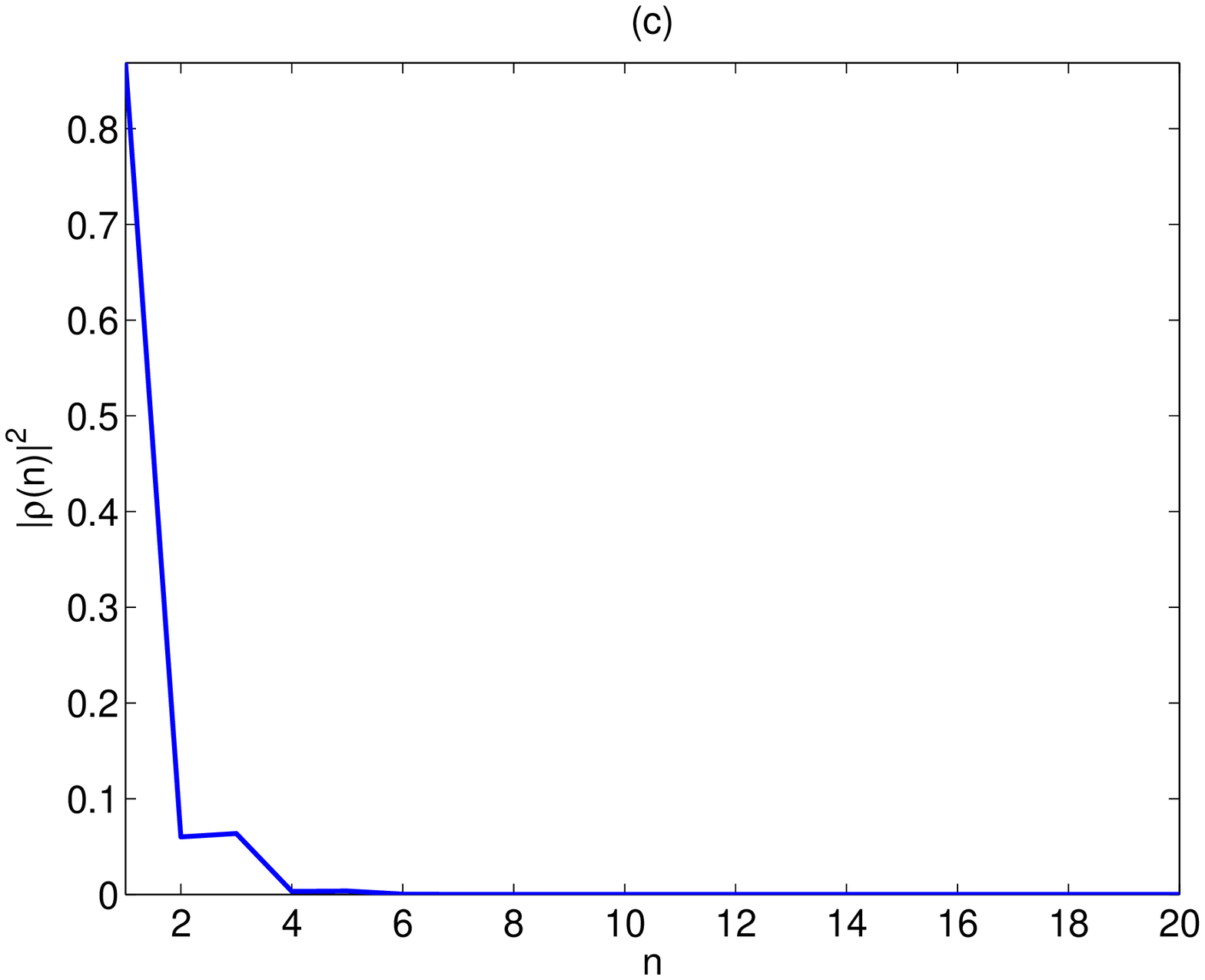} & \includegraphics[width=0.49 \columnwidth]{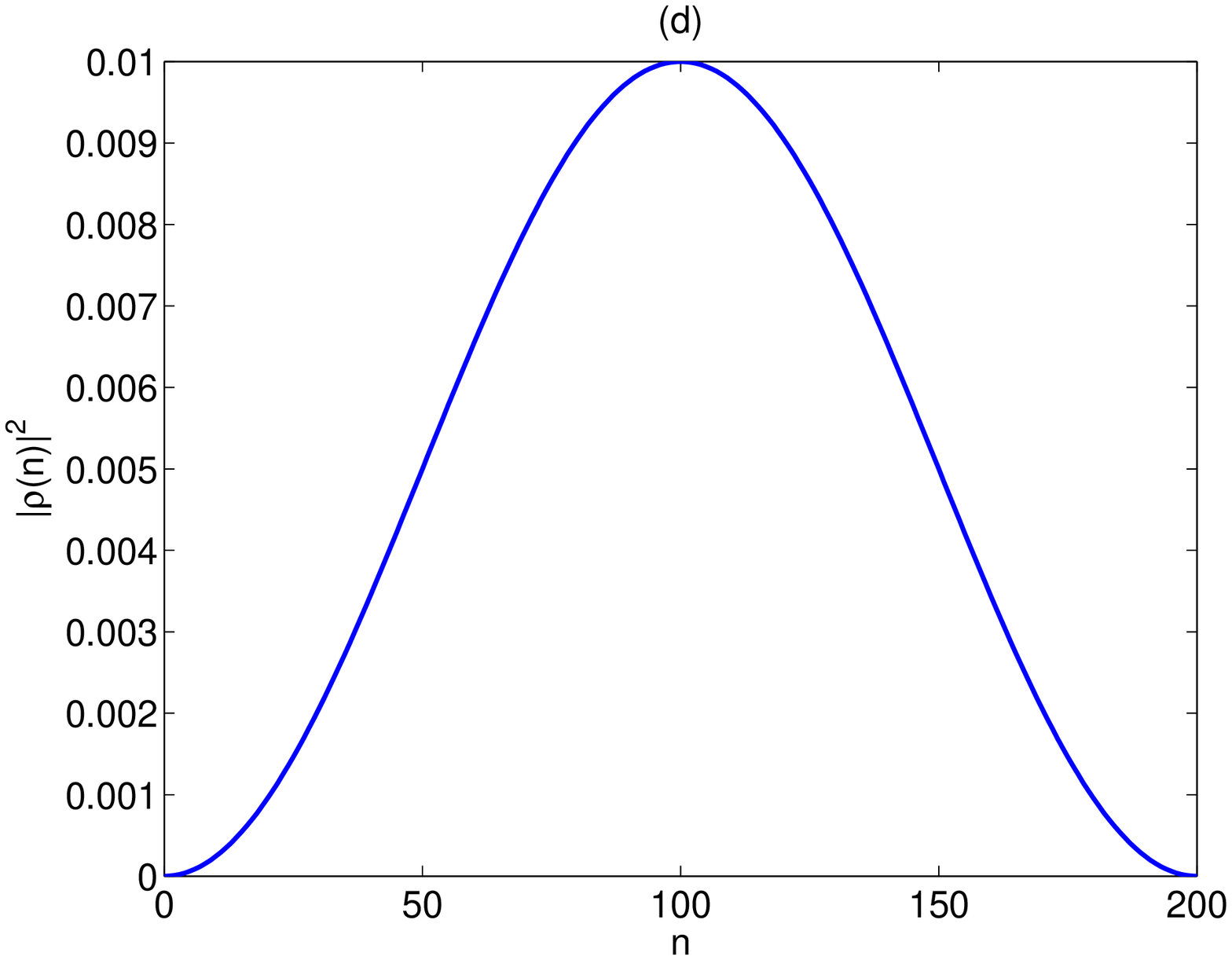}
\end{tabular}
\caption{\label{Fig3}
(Color online) Occupation probabilities $|\rho(n)|^2$ as a function of unit cell index $n$ of selected edge (a-c) and bulk (d) states for $J_{1,x}=1.0$ and $J_{2,x}=0.5$ with different values of
$J_{1,y}$ and $J_{2,y}$. In figure (a) $J_{1,y}=0.2$, $J_{2,y}=0.1$, and $k_y=0$ in figure (b) $J_{1,y}=1.0$, $J_{2,y}=0.7$, and $k_y=\pi$,
in figure (c) $J_{1,y}=0.8$, $J_{2,y}=0.1$, and $k_y=0$, and in figure (d) $J_{1,y}=1.5$, $J_{2,y}=0.8$, and $k_y=0$.
}
\end{figure}

In order to confirm these analytical results based on the winding number
given in Ref.~\cite{Matsuura}, we numerically determined all eigenvalues
by exact diagonalization of Hamiltonian (\ref{eq:Ham_2D_1}) on a finite
lattice. We use periodical boundary conditions in $y$-direction and
open boundary conditions in $x$-direction.
Results of the numerical exact diagonalization on a $200\times 200$ lattice are shown  in figure \ref{Fig2}, which
presents the energy spectra as a function of momentum $k_y$ parallel to the
surface for $J_{1,x}=1.0$ and $J_{2,x}=0.5$ with different values of
$J_{1,y}$ and $J_{2,y}$. The values we have chosen correspond to the four
different topological phases mentioned above.
Figure \ref{Fig2}~(a) demonstrates that for $a_y<1$ and $b_y<1$, there exists
an edge flat band for all momenta $k_y$, and the bulk is an insulator.
Figure \ref{Fig2}~(b) demonstrates that for $a_y>1$ and $b_y<1$, there exists 
an edge flat band for a finite range of momenta with
\[
|k_y|>2\arccos\left(\sqrt{\frac{1-b_y^2}{a_y^2-b_y^2}}\right).
\]
Figure \ref{Fig2}~(c) demonstrates that for $a_y<1$ and $b_y>1$, there exists 
an edge flat band for a finite range of momenta with
\[
|k_y|<2\arccos\left(\sqrt{\frac{b_y^2-1}{b_y^2-a_y^2}}\right).
\]
Finally, figure \ref{Fig2}~(d) shows that for $a_y>1$ and $b_y>1$, there is
no flat band and the bulk is insulating, corresponding to a trivial insulator.

Figure \ref{Fig3} demonstrates that the wave functions of the flat band states
are indeed localized at the edge of the system.
The figure shows the occupation probabilities $|\rho(n)|^2$ as a function of unit cell index $n$ of selected edge~(a-c) and bulk~(d) states for $J_{1,x}=1.0$ and $J_{1,x}=0.5$ with different values of
$J_{1,y}$ and $J_{2,y}$. We see from figures \ref{Fig3}~(a-c) that the edge
states are well localized on the boundary within the first 5 to 10 lattice
sites. Thus the flat bands should be
visible even for small lattice size.
The localization becomes better, if $|J_{1,x}-J_{2,x}|/(J_{1,x}+J_{2,x})$ increases.

If $J_{2,x}=J_{1,x}$ we have no flat bands. In this case the analytical method described
above cannot always be used, because the integral~\eqref{eq:sch_integral_2D_1} does not converge in this case.. However, in this case we can deduce that,
if $k_y\neq 0$ and $J_{y,1}\neq J_{y,2}$, $\max(|w_{+}|,|w_{-}|)>1$ and
$\min(|w_{+}|,|w_{-}|)<1$. Thus no flat bands appear.
If $k_y=0$ or $J_{y,1}=J_{y,2}$, the integral \eqref{eq:sch_integral_2D_1} does not converge, because of poles on the integration path. In this case the projection
of the Fermi surface onto the boundary forms a continuum with the bulk states, 
thus we have no edge flat bands.

\subsection{Three-dimensional model}

The three dimensional case is a direct generalization of the 2D case.
The Hamiltonian can be written as
\beq
\label{eq:Ham_3D_1}
\begin{split}
\hat H_{3D}&=\sum_{m,n,i}\bigg[-J_{1,x}\hat a_{1,m,n,i}^{\dagger}\hat b_{1,m,n,i}-J_{2,x}\hat a_{1,m,n,i}^{\dagger}\hat b_{1,m-1,n,i}
-J_{1,x}\hat a_{2,m,n,i}^{\dagger}\hat b_{2,m+1,n,i}-J_{2,x}\hat a_{2,m,n,i}^{\dagger}\hat b_{2,m,n,i}\\
&-J_{1,x}\hat a_{3,m,n,i}^{\dagger}\hat b_{3,m+1,n,i}-J_{2,x}\hat a_{3,m,n,i}^{\dagger}\hat b_{3,m,n,i}
-J_{1,x}\hat a_{4,m,n,i}^{\dagger}\hat b_{4,m,n,i}-J_{2,x}\hat a_{4,m,n,i}^{\dagger}\hat b_{4,m-1,n,i}\\
&-J_{1,y}\hat a_{1,m,n,i}^{\dagger}\hat b_{2,m,n,i}-J_{2,y}\hat a_{1,m,n,i}^{\dagger}\hat b_{2,m,n-1,i}
-J_{1,y}\hat a_{2,m,n,i}^{\dagger}\hat b_{1,m,n+1}-J_{2,y}\hat a_{2,m,n,i}^{\dagger}\hat b_{1,m,n,i}\\
&-J_{1,y}\hat a_{3,m,n,i}^{\dagger}\hat b_{4,m,n+1,i}-J_{2,y}\hat a_{3,m,n,i}^{\dagger}\hat b_{4,m,n,i}
-J_{1,y}\hat a_{4,m,n,i}^{\dagger}\hat b_{3,m,n}-J_{2,y}\hat a_{4,m,n,i}^{\dagger}\hat b_{3,m,n-1,i}\\
&-J_{1,z}\hat a_{1,m,n,i}^{\dagger}\hat b_{3,m,n,i}-J_{2,z}\hat a_{1,m,n,i}^{\dagger}\hat b_{3,m,n,i-1}
-J_{1,z}\hat a_{2,m,n,i}^{\dagger}\hat b_{4,m,n,i}-J_{2,z}\hat a_{2,m,n,i}^{\dagger}\hat b_{4,m,n,i-1}\\
&-J_{1,z}\hat a_{3,m,n,i}^{\dagger}\hat b_{2,m,n,i+1}-J_{2,z}\hat a_{3,m,n,i}^{\dagger}\hat b_{2,m,n,i}
-J_{1,z}\hat a_{4,m,n,i}^{\dagger}\hat b_{1,m,n,i+1}-J_{2,z}\hat a_{4,m,n,i}^{\dagger}\hat b_{1,m,n,i}
+h.c.\bigg],
\end{split}
\enq
where $m$, $n$, and $i$ are the unit cell indices, and $\hat a$ and $\hat b$ are annihilation operators. This Hamiltonian
consists of eight sublattice sites per unit cell.

If one takes a Fourier transform of the Hamiltonian, the Hamiltonian can be written as
\beq
\label{eq:Ham_3D_2}
H_{3D}=\sum_{\bf k}\hat \Psi^{\dagger}_{\bf k}
\begin{pmatrix}
0 & 0 & 0 & 0 & f^*_x(k_x) & f^*_y(k_y) & f^*_z(k_z) & 0\\
0 & 0 & 0 & 0 & f^*_y(k_y) & f^*_x(k_x) & 0 & f^*_z(k_z)\\
0 & 0 & 0 & 0 & f^*_z(k_z) & 0 & f^*_x(k_x) & f^*_y(k_y)\\
0 & 0 & 0 & 0 & 0 & f^*_z(k_z) & f^*_y(k_y) & f^*_x(k_x)\\
f_x(k_x) & f_y(k_y) & f_z(k_z) & 0 & 0 & 0 & 0 & 0\\
f_y(k_y) & f_x(k_x) & 0 & f_z(k_z) & 0 & 0 & 0 & 0\\
f_z(k_z) & 0 & f_x(k_x) & f_y(k_y) & 0 & 0 & 0 & 0\\
0 & f_z(k_z) & f_y(k_y) & f_x(k_x) & 0 & 0 & 0 & 0\\
\end{pmatrix}\hat \Psi_{\bf k},
\enq
where ${\bf k}=(k_x,k_y,k_z)$,
\[
\hat \Psi_{\bf k}=(\hat a_{1,\bf k},\hat a_{2,\bf k},\hat a_{3,\bf k},\hat a_{4,\bf k}
,\hat b_{1,\bf k},\hat b_{2,\bf k},\hat b_{3,\bf k},\hat b_{4,\bf k})^T,
\]
and
\beq
\label{eq:k_functions_3D}
f_{\alpha}(k_{\alpha})=-J_{1,\alpha}e^{ik_{\alpha}d/2}-J_{2,\alpha}e^{-ik_{\alpha}d/2}.
\enq
As in the 2D case this Hamiltonian has a particle-hole symmetry, time-reversal
symmetry, and a parity symmetry. Most importantly the Hamiltonian 
exhibits a chiral symmetry, again, allowing for the existence of topological edge flat bands~\cite{Schnyderpt,Ryu}.
Also this Hamiltonian has an off-diagonal block form, thus the methods from reference~\cite{Schnyder,Matsuura} can
be used to treat the Hamiltonian, again. In the present case the block is given by
\beq
\label{eq:Block_3d}
D_{3D}({\bf k})=
\begin{pmatrix}
f_x(k_x) & f_y(k_y) & f_z(k_z) & 0\\ 
f_y(k_y) & f_x(k_x) & 0 & f_z(k_z)\\
f_z(k_z) & 0 & f_x(k_x) & f_y(k_y)\\
0 & f_z(k_z) & f_y(k_y) & f_x(k_x)\\
\end{pmatrix}.
\enq

If we consider a boundary perpendicular to the $x$-direction, the existence of
edge flat bands is connected to the value of the following winding number~\cite{Schnyder,Matsuura}
\beq
\label{eq:sch_integral_3D_1}
w(k_y,k_z)=\frac{1}{2\pi}\text{Im}\left(\int_{-\pi}^{\pi}\,dk_x\, \frac{\partial_{k_x}\text{Det}(D_{3D}({\bf k}))}
{\text{Det}(D_{3D}({\bf k}))}\right).
\enq
Again, this integral can be mapped to a path integral over the unit circle,
and the integrand is then given by
\beq
\label{eq:sch_integrand_3D_1}
f_{3D}(z)=\frac{p_1(z)}{zp_2(z)},
\enq
where
\[
\begin{split}
p_1(z)&=2J_{1,x}^4z^4+4J_{1,x}^3J_{2,x}z^3-4J_{2,x}^3J_{1,x}z-2J_{2,x}^4-2(J_{1,x}^2z^3-J_{2,x}^2z)(f_y(k_y)^2+f_z(k_z)^2),\\
p_2(z)&=(J_{1,x}z+J_{2,x})^4-2z[(J_{1,x}z+J_{2,x})^2(f_y(k_y)^2+f_z(k_z)^2)]+z^2[f_y(k_y)^4+f_z(k_z)^4-2f_y(k_y)^2f_z(k_z)^2].
\end{split}
\]
Integrand \eqref{eq:sch_integrand_3D_1} can be simplified as
\beq
\label{eq:sch_integrand_3D_2}
f_{3D}(z)=\frac{-2}{z}+\frac{1}{z-w^+_{+}}+\frac{1}{z-w^+_{-}}+\frac{1}{z-w^-_{+}}+\frac{1}{z-w^-_{-}},
\enq
where
\beq
\label{eq:Roots_3D}
\begin{split}
w^+_{\pm}&=\left(\frac{(f_y(k_y)+f_z(k_z))\pm\sqrt{(f_y(k_y)+f_z(k_z))^2-4J_{1,x}J_{2,x}}}{2J_{1,x}}\right)^2,\\
w^-_{\pm}&=\left(\frac{(f_y(k_y)-f_z(k_z))\pm\sqrt{(f_y(k_y)-f_z(k_z))^2-4J_{1,x}J_{2,x}}}{2J_{1,x}}\right)^2.
\end{split}
\enq
Thus the integral can assume the values $-2,-1,0,1,2$. If the value is $0$
there are no flat bands, but otherwise there are.
If the value equals $2$ (or $-2$) we find two degenerate flat bands.

Let us assume $J_{1,x}>J_{2,x}$. Now $\min(|w^+_{+}|,|w^+_{-}|)\leq J_{2,x}/J_{1,x}<1$ and $\min(|w^-_{+}|,|w^-_{-}|)\leq J_{2,x}/J_{1,x}<1$. Thus 
we have always at least two poles within the unit disc. Thus the value of the
integral depends on the values of $\max(|w^+_{+}|,|w^+_{-}|)$ and $\max(|w^-_{+}|,|w^-_{-}|)$.
It turns out that
\beq
\label{eq:max_a_value_3D}
\begin{split}
&\max(|w^{\pm}_{+}|,|w^{\pm}_{-}|)\leq \left(\frac{(J_{1,x}-J_{2,x})|V^{\pm}_{k_y,k_z}|+\sqrt{[(J_{1,x}-J_{2,x})|V^{\pm}_{k_y,k_z}|]^2+4J_{1,x}J_{2,x}}}{2J_{1,x}}\right)^2
,\text{if}\,\, |V^{\pm}_{k_y,k_z}|<1\\
&\max(|w^{\pm}_{+}|,|w^{\pm}_{-}|)=1,\hspace{276pt} \text{if}\,\, |V^{\pm}_{k_y,k_z}|=1\\
&\max(|w^{\pm}_{+}|,|w^{\pm}_{-}|)\geq \left(\frac{(J_{1,x}-J_{2,x})|V^{\pm}_{k_y,k_z}|+\sqrt{[(J_{1,x}-J_{2,x})|V^{\pm}_{k_y,k_z}|]^2+4J_{1,x}J_{2,x}}}{2J_{1,x}}\right)^2
,\text{if}\,\, |V^{\pm}_{k_y,k_z}|> 1,
\end{split}
\enq
where
\[
\begin{split}
|V^{\pm}_{k_y,k_z}|^2
&=\left(\frac{J_{1,y}+J_{2,y}}{J_{1,x}+J_{2,x}}\cos(k_y/2)\pm \frac{J_{1,z}+J_{2,z}}{J_{1,x}+J_{2,x}}\cos(k_z/2)\right)^2
+\left(\frac{J_{1,y}-J_{2,y}}{J_{1,x}-J_{2,x}}\sin(k_y/2)\pm \frac{J_{1,z}-J_{2,z}}{J_{1,x}-J_{2,x}}\sin(k_z/2)\right)^2\\
&=(a_y\cos(k_y/2)\pm a_z\cos(k_z/2))^2+(b_y\sin(k_y/2)\pm b_z\sin(k_z/2))^2.
\end{split}
\]
If $|V^{\pm}_{k_y,k_z}|<1$, $\max(|w^{\pm}_{+}|,|w^{\pm}_{-}|)<1$, and if $|V^{\pm}_{k_y,k_z}|>1$, $\max(|w^{\pm}_{+}|,|w^{\pm}_{-}|)>1$.
If both $\max(|w^{\pm}_{+}|,|w^{\pm}_{-}|)$ are smaller than 1 we find two
flat bands, if only one is smaller than 1 we find a single flat band, if 
both are larger than 1 we find no flat bands.

The location (in the projected momentum space, i.e. in the $k_yk_z$-plane) of
the flat bands depends on the parameters $a_y,a_z,b_y,b_z$. Without loss of generality we can assume all these parameters non-negative. In this case $\sup_{k_y,k_z}|V^+_{k_y,k_z}|\geq \sup_{k_y,k_z}|V^-_{k_y,k_z}|$ and $\inf_{k_y,k_z}|V^+_{k_y,k_z}|\geq \inf_{k_y,k_z}|V^-_{k_y,k_z}|$. The values of $\sup_{k_y,k_z}|V^+_{k_y,k_z}|$, $\sup_{k_y,k_z}|V^-_{k_y,k_z}|$, $\inf_{k_y,k_z}|V^+_{k_y,k_z}|$, and $\inf_{k_y,k_z}|V^-_{k_y,k_z}|$ can be found in Appendix B.

We have several different phase possibilities: (I) if
$\sup_{k_y,k_z}|V^+_{k_y,k_z}|<1$, the bulk is an insulator, and we find two
isolated flat bands for every $\tilde {\bf k}=(k_y,k_z)$. (II) If $\sup_{k_y,k_z}|V^-_{k_y,k_z}|<1$, $\sup_{k_y,k_z}|V^+_{k_y,k_z}|\geq 1$, and
$\inf_{k_y,k_z}|V^+_{k_y,k_z}|<1$, the bulk is a semimetal we find a flat band
for all $\tilde {\bf k}$. For some  $\tilde {\bf k}$
there are two flat bands, and one flat band for the rest. (III) If $\inf_{k_y,k_z}|V^+_{k_y,k_z}|<1$ and $\sup_{k_y,k_z}|V^-_{k_y,k_z}|\geq 1$,
the bulk is a semimetal. Flat bands can be found only in some part of the projected $k$-space.
(IV) If $\inf_{k_y,k_z}|V^+_{k_y,k_z}|>1$, $\inf_{k_y,k_z}|V^-_{k_y,k_z}|<1$, and $\sup_{k_y,k_z}|V^-_{k_y,k_z}|\geq 1$, the bulk is a semimetal, and we have a non-degenerate flat band in some part of the projected momentum space.
(V) If $\inf_{k_y,k_z}|V^-_{k_y,k_z}|>1$, the bulk is an insulator, and we find no flat bands.

One could move from one phase to another by tuning the laser intensity.

\begin{figure}[t]
\begin{tabular}{ll}
 \includegraphics[width=0.49 \columnwidth,angle=270]{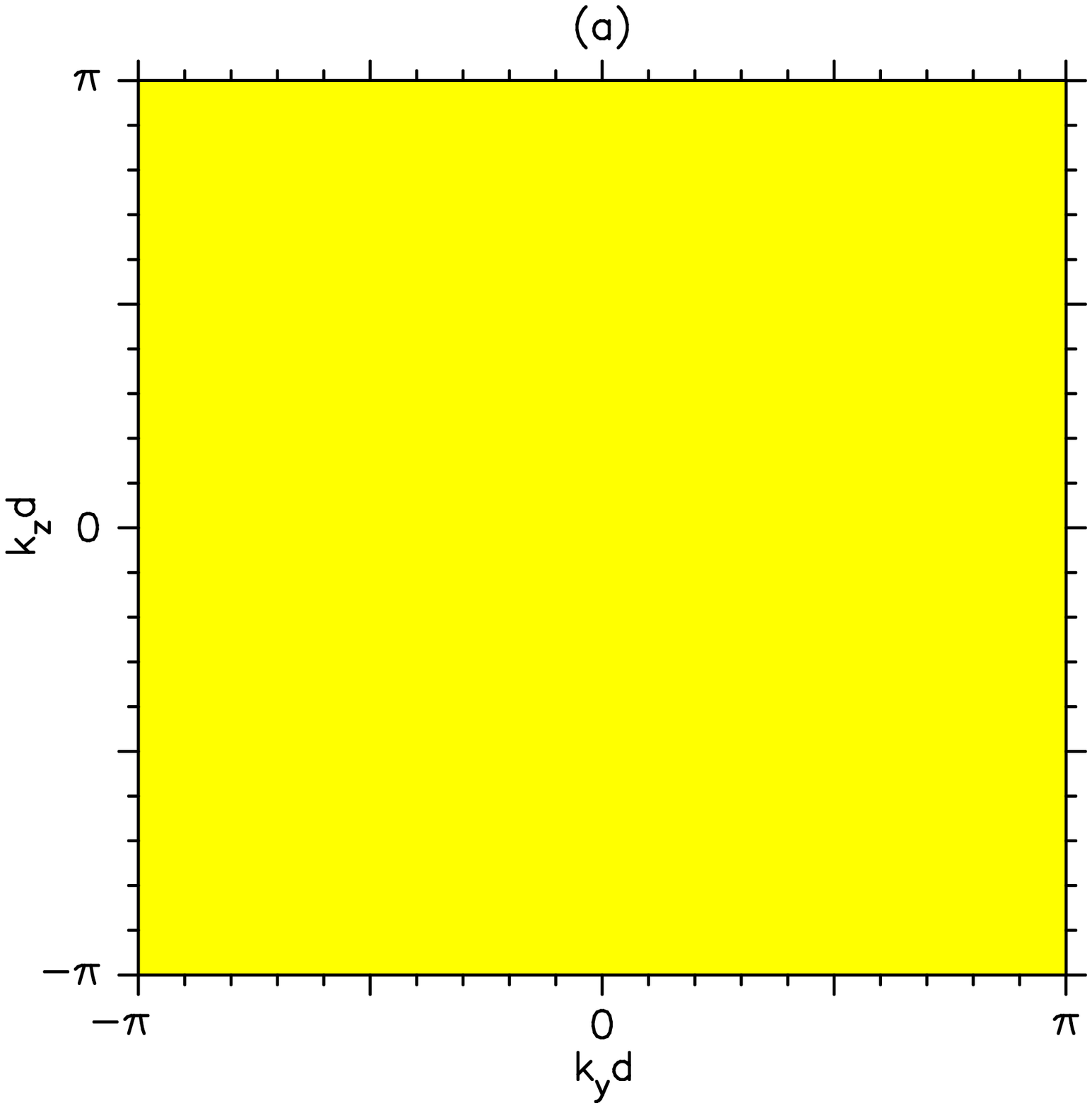} & \includegraphics[width=0.49 \columnwidth,angle=270]{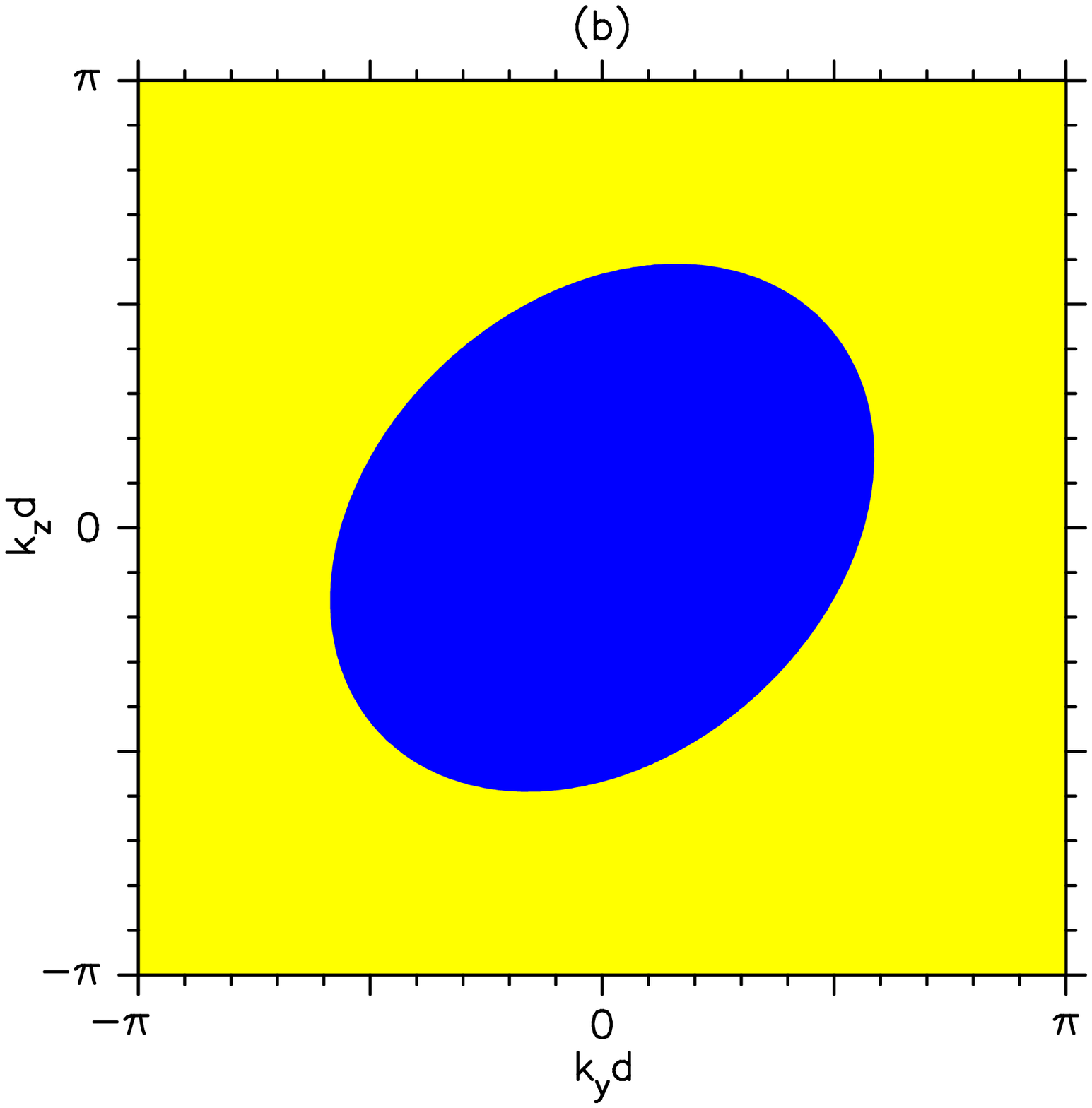}\\
 \includegraphics[width=0.49 \columnwidth,angle=270]{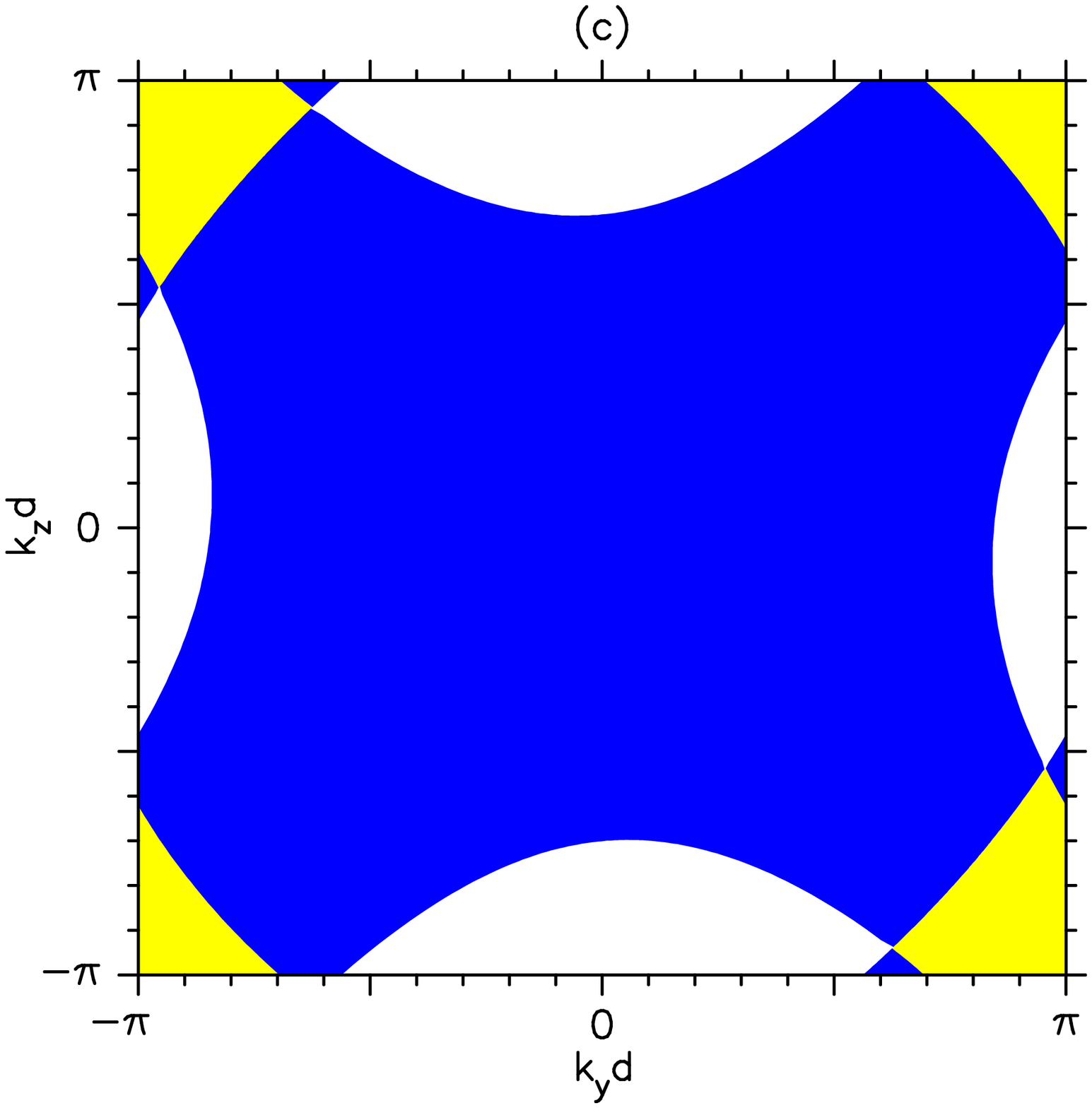} & \includegraphics[width=0.49 \columnwidth,angle=270]{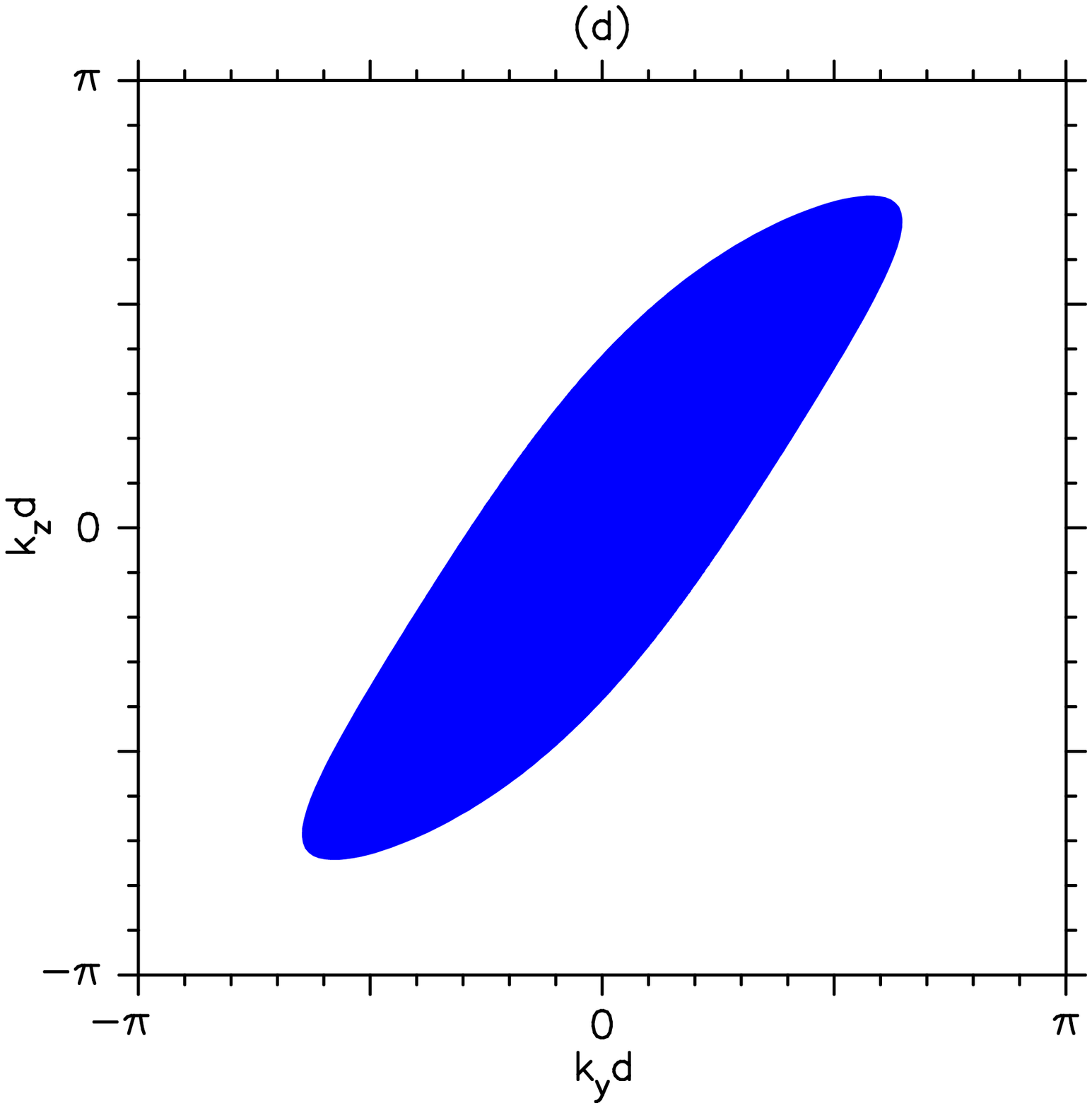}
\end{tabular}
\caption{\label{Fig4}
(Color online) Examples of 2D flat bands with different phases in the 3D
  case. Yellow (light gray): a two-fold degenerate zero energy flat band.
Blue (dark gray): a non-degenerate zero energy flat band. White: no flat bands.
 Figure~(a) phase 1 $J_{1,y}=J_{1,z}=0.2$ and $J_{2,y}=J_{2,z}=0.1$.
Figure~(b) phase 2 $J_{1,y}=0.6$, $J_{1,z}=0.4$, and $J_{2,y}=J_{2,z}=0.2$.
Figure~(c) phase 3 $J_{1,y}=J_{1,z}=1.0$, $J_{2,y}=0.8$, and $J_{2,z}=0.6$.
Figure~(d) phase 4 $J_{1,y}=2.0$, $J_{1,z}=1.2$, $J_{2,y}=0.3$, and $J_{2,z}=0.4$.
In all these  figures $J_{1,x}=1.0$ and $J_{2,x}=0.2$.}
\end{figure}

Figure~\ref{Fig4} shows some examples of 2D flat bands in the surface
Brillouin zone for the different phases.
In figure~\ref{Fig4}~(a) we show phase I with $J_{1,y}=J_{1,z}=0.2$ and $J_{2,y}=J_{2,z}=0.1$, here we see that the two-fold degenerate
flat band totally fills the projected momentum space.
Figure~\ref{Fig4}~(b) shows phase II with $J_{1,y}=0.6$, $J_{1,z}=0.4$, and
$J_{2,y}=J_{2,z}=0.2$, the two-fold degenerate flat band
fills the projected momentum space only partially, and the rest of the momentum space (near the origin) is filled by a non-degenerate flat band. 
Phase III with $J_{1,y}=J_{1,z}=1.0$, $J_{2,y}=0.8$, and $J_{2,z}=0.6$ is shown in figure~\ref{Fig4}~(c). In this case we see that only a part of the projected momentum space is occupied by the flat bands (degenerate and non-degenerate).
In figure~\ref{Fig4}~(d) we show phase IV with $J_{1,y}=2.0$, $J_{1,z}=1.2$,
$J_{2,y}=0.3$, and $J_{2,z}=0.4$. Here we see that only a non-degenerate flat
band partially fills the projected momentum space.
Throughout figure~\ref{Fig4} we have set $J_{1,x}=1.0$ and $J_{1,x}=0.2$.

If $J_{1,x}=J_{2,x}$ there are no flat bands. The reasons are the same as in the 2D-case.
If one looks at all three possible boundaries (perpendicular to $x$-, $y$-, or
$z$-direction), 
the only situation when one cannot find any flat bands
is the case, where $J_{1,x}=J_{2,x}$, $J_{1,y}=J_{2,y}$, and $J_{1,z}=J_{2,z}$. 
In this case the bulk is a metal (a standard cubic lattice).
Thus if the bulk is an insulator or a semimetal one can always find flat bands
at least at one of the boundaries.

\section{Conclusion}

We have presented both 2D and 3D optical lattices, which possess topological
surface flat bands. We have proven the unusual topological properties by
two independent methods: an analytical calculation based on a topological
winding number and numerical calculations of the eigenstates of the systems.
These lattices can be relatively easy created by counterpropagating laser
beams and do not require spin-orbit coupling nor non-Abelian gauge fields.
By tuning the intensity of the potential it is possible to sweep between
topological insulating, semi-metallic, and trivial insulating phases.

The surface flat bands found here are protected by the particle-hole
chiral symmetry given in Eq.~(\ref{eq:S}), which requires the
energy spectrum to be symmetric with respect to energy
$E=0$. In contrast to solid state systems 
optical lattices can be realized experimentally with high 
precision to follow this symmetry. Even in the case that a 
small particle-hole breaking term is present in the Hamiltonian,
the flat band is not destroyed, but adiabatically deformed
and becomes slightly dispersive \cite{Paananen_1,Paananen_2}.

We derived simple analytical formulas for the existence and the location of the flat bands.
We showed that in the 3D case we can have a double degeneracy of the flat
bands. Also in the 3D case the flat bands are two dimensional,
in contrast to other surface flat band systems, where the flat bands are one
dimensional. Due to their short localization length, these flat bands could be 
realized in experiments even with a small lattice size.

These systems can be used to study the influence of interaction on the
flat bands via a Feshbach resonance or to study high-$T_c$ surface superfluidity. Also
these systems can be used to model different flat band systems with little modification.

\appendix

\section{Creation of the lattices}
\label{App:App_I}

\begin{figure}[t]
 \includegraphics[width=0.98 \columnwidth]{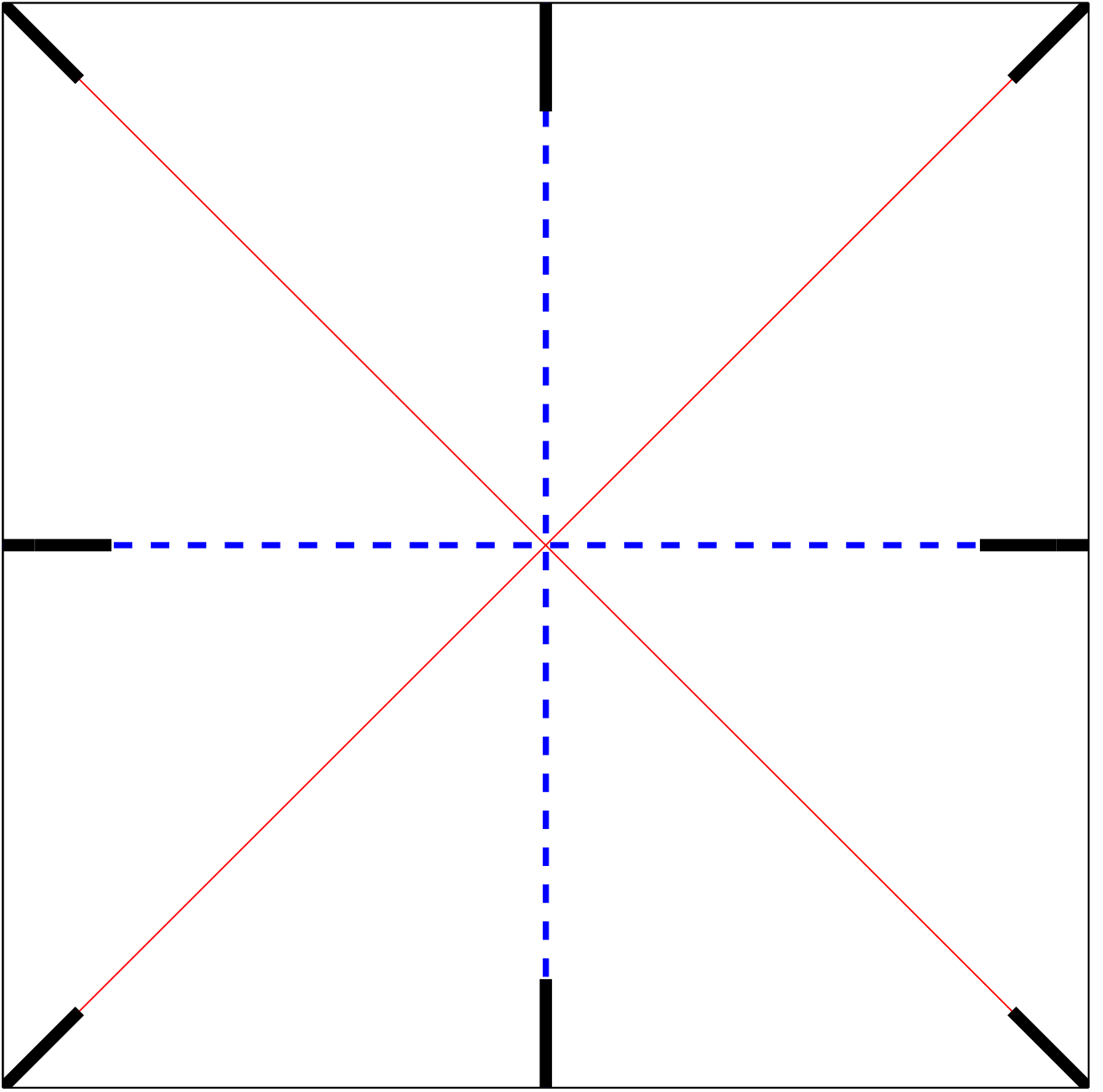} 
\caption{\label{Fig5} (Color online) Schematic figure of the laser
  arrangement. Black sticks indicate the lasers, red solid lines indicate
  $\sqrt{2}\lambda$-laser beams, and blue dashed lines $\lambda$-laser beams.}
\end{figure}

\begin{figure}[t]
 \includegraphics[width=0.98 \columnwidth]{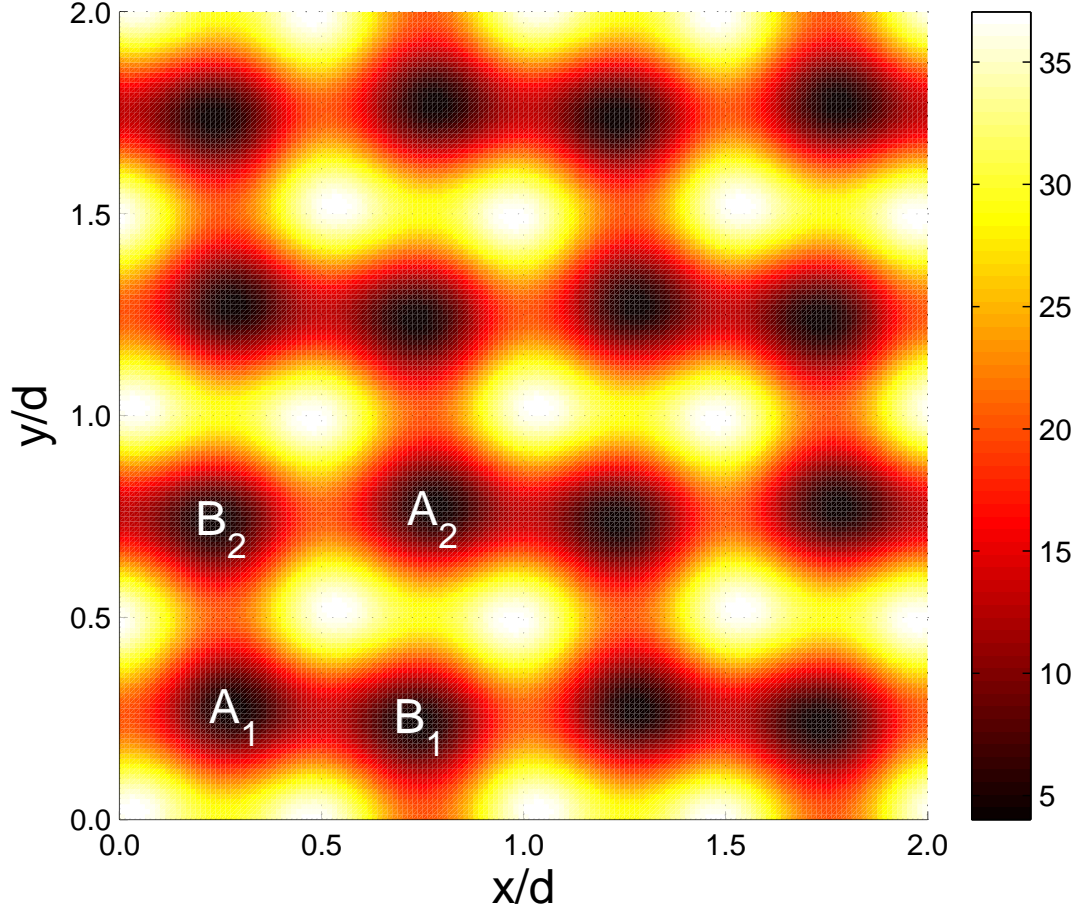} 
\caption{\label{Fig6} (Color online) 2D lattice potential with $V_1=12E_r$, $V_2=20E_r$, $V_3=8E_r$, $V_4=1E_r$, $\theta_1=\pi/4$, and $\theta_2=-\pi/4$, $E_r=\hbar^2 \pi^2/(md^2)$ is the
recoil energy. The color coding is such as: white=high values,
black=low
values. The four lattice sites $A_1$, $A_2$, $B_1$, and $B_2$ from Fig.~\ref{Fig1}
are indicated in the first unit cell, for comparison.}
\end{figure}

The corresponding 2D lattice can be formed using the following periodical potential
\beq
\label{eq:2D_lat_pot}
V_{2D}(x,y)=V_1\cos^2\left(\frac{2\pi x}{d}\right)+V_2\cos^2\left(\frac{2\pi y}{d}\right)
+V_3\cos^2\left(\frac{\pi(x+y)}{d}+\theta_1\right)+V_4\cos^2\left(\frac{\pi(x-y)}{d}+\theta_2\right).
\enq
To create this potential we need four lasers with wavelength $\lambda_1=d$ and four lasers with
wavelength $\lambda_2=\sqrt{2}d$, where $d$ is the periodicity of the lattice. 
Figure~\ref{Fig5} shows the laser arrangement schematically.
We note that the four terms in Eq.~(\ref{eq:2D_lat_pot}) should not interfere
  with each other. Experimentally, this can be accomplished by choosing the field directions
  perpendicular to each other or control of the relative time-phase delay \cite{Hemmerich}.
Figure~\ref{Fig6} shows an example of the 2D lattice potential. We can see
from the figure that the barriers between the sublattice sites are different
in the forward direction (alongside the axis) than in backward direction. 
This means that also the hopping strengths are different.
As regards the quality of the edges of the system we note that sharp
edges are not a critical requirement
for existence of topological surface states. The existence of the
surface flat bands is topologically protected by the chiral symmetry
Eq.~(\ref{eq:S}).
This symmetry remains valid also for smooth edges. 

The three dimensional lattice is a generalization of the 2D case. There are
several ways to construct the three dimensional potential.
For the following potential one needs 14 lasers, six with wavelength $\lambda_1$ and eight with $\lambda_2$.
The potential is written as
\beq
\label{eq:3D_lat_pot_1}
\begin{split}
V_{3D,1}(x,y,z)&=V_1\cos^2\left(\frac{2\pi x}{d}\right)+V_2\cos^2\left(\frac{2\pi y}{d}\right)+V_3\cos^2\left(\frac{2\pi z}{d}\right)
+V_4\cos^2\left(\frac{\pi(x+y)}{d}+\theta_1\right)\\
&+V_5\cos^2\left(\frac{\pi(x-y)}{d}+\theta_2\right)+V_6\cos^2\left(\frac{\pi(x+z)}{d}+\theta_3\right)+V_7\cos^2\left(\frac{\pi(x-z)}{d}+\theta_4\right).
\end{split}
\enq
Alternatively, a more symmetric potential is given by
\beq
\label{eq:3D_lat_pot_2}
\begin{split}
V_{3D,2}(x,y,z)&=V_1\cos^2\left(\frac{2\pi x}{d}\right)+V_2\cos^2\left(\frac{2\pi y}{d}\right)+V_3\cos^2\left(\frac{2\pi z}{d}\right)
+V_4\cos^2\left(\frac{\pi(x+y)}{d}+\theta_1\right)\\
&+V_5\cos^2\left(\frac{\pi(x-y)}{d}+\theta_2\right)+V_6\cos^2\left(\frac{\pi(x+z)}{d}+\theta_3\right)+V_7\cos^2\left(\frac{\pi(x-z)}{d}+\theta_4\right)\\
&+V_8\cos^2\left(\frac{\pi(y+z)}{d}+\theta_5\right)+V_9\cos^2\left(\frac{\pi(y-z)}{d}+\theta_6\right).
\end{split}
\enq
However, 18 lasers are needed to construct this potential.

\section{Suprema and infima for the 3D lattice}
\label{App:App_II}

If $J_{1,x}\neq J_{2,x}$ infima and suprema of $|V^+_{k_y,k_z}|$ and $|V^-_{k_y,k_z}|$ over the projected momentum space can be presented as functions of $a_y,a_z,b_y,b_z$. Without losing generality, we can assume $a_y,a_z,b_y,b_z\geq 0$. The supremum of $|V^+_{k_y,k_z}|$ is given by
\beq
\label{eq:sup_V_p_1}
\sup_{k_y,k_z}|V^+_{k_y,k_z}|
=\begin{cases}
a_y+a_z, &\text{if}\, a_y\geq b_y,\, a_z\geq b_z\\
b_y+b_z, &\text{if}\, a_y\leq b_y,\, a_z\leq b_z\\
\sqrt{\frac{(a_y^2b_z^2-a_z^2b_y^2)(a_y^2-b_y^2+b_z^2-a_z^2)}{(a_y^2-b_y^2)(b_z^2-a_z^2)}}, &\text{if}\, (a_y-b_y)(b_z-a_z)>0,\, 
\text{and}\, x_0,y_0\in \mathbb{R},\, \text{and}\,|x_0|,|y_0|\leq 1\\
\max(a_y+a_z,b_y+b_z), &\text{otherwise}
\end{cases},
\enq
where
\[
\begin{split}
x_0&=\frac{b_y}{a_y^2-b_y^2}\sqrt{\frac{-(a_y^4a_z^2+a_z^2b_y^4-a_y^2[a_z^4+b_z^4+2a_z^2(b_y^2-b_z^2)])}{a_y^2b_z^2-a_z^2b_y^2}}\\
y_0&=\frac{b_z}{b_z^2-a_z^2}\sqrt{\frac{-(a_y^4a_z^2+a_z^2b_y^4-a_y^2[a_z^4+b_z^4+2a_z^2(b_y^2-b_z^2)])}{a_y^2b_z^2-a_z^2b_y^2}}.
\end{split}
\]
The infimum of $|V^+_{k_y,k_z}|$ is given by
\beq
\label{eq:inf_V_p_1}
\inf_{k_y,k_z}|V^+_{k_y,k_z}|
=\begin{cases}
|b_y-b_z|, &\text{if}\, a_y\geq b_y,\, a_z\geq b_z\\
a_y\sqrt{1-\frac{b_z^2}{b_y^2-a_y^2}}, &\text{if}\,  a_y<b_y,\, \frac{b_yb_z}{b_y^2-a_y^2}\leq 1\\
a_z\sqrt{1-\frac{b_y^2}{b_z^2-a_z^2}}, &\text{if}\,  a_z<b_z,\, \frac{b_yb_z}{b_z^2-a_z^2}\leq 1\\
|b_y-b_z|, &\text{otherwise}
\end{cases}.
\enq
The supremum of $|V^-_{k_y,k_z}|$ is given by
\beq
\label{eq:sup_V_m_1}
\sup_{k_y,k_z}|V^-_{k_y,k_z}|
=\begin{cases}
\max\left(a_y,\sqrt{a_z^2+b_y^2}\right), &\text{if}\, b_z=0\\
\max\left(a_z,\sqrt{a_y^2+b_z^2}\right), &\text{if}\, b_y=0\\
b_y+b_z, &\text{if}\, a_y\leq b_y,\, a_z\leq b_z\\
a_y\sqrt{1+\frac{b_z^2}{a_y^2-b_y^2}}, &\text{if}\,  a_y>b_y,\, \frac{b_yb_z}{a_y^2-b_y^2}\leq 1\,\text{and}\, a_z\leq b_z\, \text{or}\,
\frac{b_yb_z}{a_z^2-b_z^2}> 1\\
a_z\sqrt{1+\frac{b_y^2}{a_z^2-b_z^2}}, &\text{if}\,  a_z>b_z,\, \frac{b_yb_z}{a_z^2-b_z^2}\leq 1\,\text{and}\, a_y\leq b_y\, \text{or}\,
\frac{b_yb_z}{a_y^2-b_y^2}> 1\\
\max\left(a_y\sqrt{1+\frac{b_z^2}{a_y^2-b_y^2}},a_z\sqrt{1+\frac{b_y^2}{a_z^2-b_z^2}}\right),&\text{if}\, a_z>b_y,\, a_z>b_z\, \text{and}\,\frac{b_yb_z}{a_y^2-b_y^2}\leq 1,\, \frac{b_yb_z}{a_z^2-b_z^2}\leq 1\\
b_y+b_z, &\text{otherwise}
\end{cases}.
\enq
The infimum of $|V^-_{k_y,k_z}|$ is given by
\beq
\label{eq:inf_V_m_1}
\inf_{k_y,k_z}|V^-_{k_y,k_z}|
=\begin{cases}
0, &\text{if}\, (a_y-a_z)(b_z-b_y)\geq 0\\
\min(|a_y-a_z|,|b_y-b_z|), &\text{if}\, (a_y-a_z)(b_z-b_y)<0\, \text{and}\, (a_y-b_y)(a_z-b_z)\leq 0\\
\sqrt{\frac{(a_y^2b_z^2-a_z^2b_y^2)(a_y^2-b_y^2+b_z^2-a_z^2)}{(a_y^2-b_y^2)(b_z^2-a_z^2)}}, &\text{if}\, (a_y-b_y)(a_z-b_z)>0,\, 
\text{and}\, x_1,y_1\in \mathbb{R},\, \text{and}\,|x_1|,|y_1|\leq 1\\
\min(|a_y-a_z|,|b_y-b_z|), &\text{otherwise}
\end{cases},
\enq
where
\[
\begin{split}
x_1&=\frac{b_y}{a_y^2-b_y^2}\sqrt{\frac{-(a_y^4a_z^2+a_z^2b_y^4-a_y^2[a_z^4+b_z^4+2a_z^2(b_y^2-b_z^2)])}{a_y^2b_z^2-a_z^2b_y^2}}\\
y_1&=\frac{b_z}{a_z^2-b_z^2}\sqrt{\frac{-(a_y^4a_z^2+a_z^2b_y^4-a_y^2[a_z^4+b_z^4+2a_z^2(b_y^2-b_z^2)])}{a_y^2b_z^2-a_z^2b_y^2}}.
\end{split}
\]

\end{document}